\documentclass[sigconf]{acmart}
\usepackage{endnotes}
\usepackage[compact]{titlesec} 
\usepackage{url}
\usepackage[skip=0.3\baselineskip]{caption}

\usepackage{breakurl}

\renewenvironment{quote}
  {\list{}{\rightmargin=0.4cm \leftmargin=0.4cm}%
   \item\relax}
  {\endlist}

\titlespacing\section{0pt}{0.6em plus 0.0em minus 0.1em}{0.2em plus 0.1em minus 0.1em}
\titlespacing\subsection{0pt}{0.6em plus 0.0em minus 0.1em}{0.2em plus 0.1em minus 0.1em}
\titlespacing\subsubsection{0pt}{0.1em plus 0.1em minus 0.0em}{0.5em plus 0.1em minus 0.1em}

\copyrightyear{2021}
\acmYear{2021}
\setcopyright{acmlicensed}\acmConference[FAccT '21]{Conference on Fairness, Accountability, and Transparency}{March 3--10, 2021}{Virtual Event, Canada}
\acmBooktitle{Conference on Fairness, Accountability, and Transparency (FAccT '21), March 3--10, 2021, Virtual Event, Canada}
\acmPrice{15.00}
\acmDOI{10.1145/3442188.3445888}
\acmISBN{978-1-4503-8309-7/21/03}

\begin{document}

\title["What We Can't Measure, We Can't Understand"]{"What We Can't Measure, We Can't Understand": Challenges to Demographic Data Procurement in the Pursuit of Fairness}


\author{McKane Andrus}
\email{mckane@partnershiponai.org}
\author{Elena Spitzer}
\email{emspitzer1@gmail.com}
\affiliation{%
  \institution{Partnership on AI}
}


\author{Jeffrey Brown}
\email{jeff@partnershiponai.org}
\affiliation{Partnership on AI}
\affiliation{Minnesota State University, Mankato}

\author{Alice Xiang}\authornote{Research funded by the Partnership on AI. Senior author listed last.}
\email{alice.xiang@sony.com}
\affiliation{Sony AI}
\affiliation{Partnership on AI}

\renewcommand{\shortauthors}{Andrus, Spitzer, Brown, and Xiang}

\begin{abstract}
As calls for fair and unbiased algorithmic systems increase, so too does the number of individuals working on algorithmic fairness in industry. However, these practitioners often do not have access to the demographic data they feel they need to detect bias in practice. Even with the growing variety of toolkits and strategies for working towards algorithmic fairness, they almost invariably require access to demographic attributes or proxies. We investigated this dilemma through semi-structured interviews with 38 practitioners and professionals either working in or adjacent to algorithmic fairness. Participants painted a complex picture of what demographic data availability and use look like on the ground, ranging from not having access to personal data of any kind to being legally required to collect and use demographic data for discrimination assessments. In many domains, demographic data collection raises a host of difficult questions, including how to balance privacy and fairness, how to define relevant social categories, how to ensure meaningful consent, and whether it is appropriate for private companies to infer someone's demographics. Our research suggests challenges that must be considered by businesses, regulators, researchers, and community groups in order to enable practitioners to address algorithmic bias in practice. Critically, we do not propose that the overall goal of future work should be to simply lower the barriers to collecting demographic data. Rather, our study surfaces a swath of normative questions about how, when, and whether this data should be procured, and, in cases where it is not, what should still be done to mitigate bias.
\end{abstract}




\maketitle

\section{Introduction} \label{sec:intro}

As the risk of algorithmic discrimination has increased in recent years, so too has the number of proposed fixes from the field of algorithmic fairness. Countless fairness strategies, metrics, and toolkits have been developed and, more rarely, integrated into algorithmic systems and products. Most of these innovations revolve around measuring and sometimes mitigating disparities in treatment and performance across a “sensitive” or “protected” attribute. Typically, this process will require access to data revealing this “sensitive” or “protected” attribute. Using open-access, technical toolkits as an example, the fairness methods in Fairlearn (Microsoft) \cite{microsoftFairlearn2020}, AI Fairness 360 (IBM) \cite{bellamyAI2018}, What-if tool (Google) \cite{wexlerWhatIf2020},  LiFT (LinkedIn) \cite{vasudevanLiFT2020}, Aequitas \cite{saleiroAequitas2019}, and FAT-forensics \cite{sokol2020fat-forensics} all require access to demographic data. In many situations, however, information about demographics can be extremely difficult for practitioners to even procure \cite{holsteinImproving2019,vealeFairer2017}. For the toolkits that do not require demographic data access, they generally rely on simulated data (e.g. \cite{fairness_gym}) or approach the problem more from the angle of design (e.g. \cite{barghoutiAlgorithmic2020}).

Demographic data stands apart from much of the other types of data fed into algorithmic systems in that its collection and use are socially and legally constrained \cite{andrusWorking2020}. Data protection regulation, such as the EU’s General Data Protection Regulation (GDPR) \cite{gdpr}, builds in extra protections for data about certain attributes, deeming it off-limits for many use cases. Perceiving this potential conflict between data availability and making systems less discriminatory, legal scholars have argued for regulatory provisions for collecting demographic data to audit algorithmic bias \cite{zliobaiteUsing2016, zarskyUnderstanding2014,xiang2020reconciling,williamsHow2018,tischbirek2020artificial}.

On the other hand, computer science researchers have proposed a number of methods and techniques that try to bypass the collection of demographic data through inference and proxies (e.g., \cite{romanovWhat2019,guptaProxy2018,zhangAssessing2016}), to only use demographic data during model training (e.g., \cite{kamishimaFairnessaware2011,zafarFairness2017,sattigeriFairness2019, ghili2019eliminating}), or to identify and mitigate algorithmic discrimination for computationally identified groups without the use of demographics at all (e.g., \cite{benthallRacial2019,hashimotoFairness2018,lahotiFairness2020}). Coming at the problem from a privacy angle, others have sought to make demographic data collection and use more private and less centralized through various combinations of data sanitization, cryptographic privacy, differential privacy, and third-party auditors (e.g., \cite{hajianDiscrimination2015,jagielski2019differentially,kilbertusBlind2018,kuppamFair2020,vealeFairer2017}).


Acknowledging that these alternative methods exist, our goal was to understand how demographic data availability is encountered and dealt with in practice. By interviewing a wide range of practitioners dealing with concerns around algorithmic fairness, we strove to characterize the actual impediments to demographic data use as they emerge on the ground. We also assessed a number of concerns practitioners have around how to ensure that their demographic data collection process is responsible and ethical.  Though there is some question as to whether algorithmic fairness techniques meaningfully address discrimination and injustice and thereby justify the collection and use of this sensitive data \cite{barabasBias2019,hoffmannWhere2019,katellSituated2020,selbstFairness2019}, understanding how practitioners in industry think about and address the use of demographic data is an important piece of this debate. 

\section{Background and Related Research} \label{sec:back}
Past work on the barriers to demographic data procurement and use has largely considered the question of legality. While it is clear that data protection regulations like GDPR are likely to inhibit the procurement, storage, and use of many demographic attributes, it is less clear what types of carve-outs might apply if using the data for fairness purposes \cite{goodman2016step}. There is also an open question of whether or not anti-discrimination law permits the inclusion of demographic attributes in decision-making processes \cite{xiang2020reconciling,bentAlgorithmic2020}. \citet{bogenAwareness2020} survey the requirements of U.S anti-discrimination law around the collection and usage of sensitive attribute data in employment, credit, and healthcare, and find that “there are few clear, generally accepted principles about when and why companies should collect sensitive attribute data for anti-discrimination purposes.” \cite[p. 2]{bogenAwareness2020}

A more critical branch of scholarship is now interrogating the algorithmic fairness community’s conceptualization of demographic attributes themselves and how that propagates into notions of fairness. Work in this area has addressed race \cite{hannaCritical2020,benthallRacial2019}, gender \cite{hamidiGender2018,scheuermanHow2020,huWhat2020}, and disability \cite{bennettWhat2020}, calling into question what it means to even rely on these categories as a basis for assessing unfairness, and what harms are reproduced by relying on these infrastructures of categorization. Similar to this work, important contributions emerging from various academic fields and activist communities on the issue of justice in data use interrogate when and where it is acceptable to collect what types of data and what degree of control data subjects should have over their data thereafter \cite{taylorWhat2017,ciforFeminist2019,milanBig2019,rainie2019indigenous,petty2018our,Data}. These lines of work center questions of autonomy and representation around data and its use, going beyond the common legal standards of privacy and anti-discrimination.

Less work has been done, however, in understanding how practitioners confront issues around demographic data procurement and use and if the difficulties they encounter mirror those discussed above. There is a growing literature of work studying fairness practitioners themselves that has focused on the needs of public sector practitioners \cite{vealeFairness2018}, the needs of private sector practitioners \cite{holsteinImproving2019} and the organizational patterns they fit within \cite{rakovaWhere2020}. These studies lay important groundwork in identifying the problem,\footnote{Holstein et al. found, for example, that a majority of their survey respondents “indicated that the availability of tools to support fairness auditing without access to demographics at an individual level would be at least ‘Very’ useful” \cite[p. 8]{holsteinImproving2019}.} but with this paper we aim to provide more texture as to how practitioners themselves think about and navigate this gap, and, in turn, what paths forward would need to address in order to be effective.

\section{Methods} \label{sec:meth}

Building off of themes from prior work in this area \cite{andrusJust2019}, we scoped an interview study looking at how demographic data procurement and use actually proceeds in practice. We interviewed 38 practitioners from 26 organizations, with 5 participants being the maximum number coming from a single organization. All participants either A) were involved in efforts to detect bias or unfairness in a model or product, B) were familiar with company policies relevant to the usage of demographic data, or C) were familiar with regulations or policies relevant to the collection or use of demographic data.  

Most participants were from for-profit tech companies (34 out of 38). A slight majority (55\%) of participants were from companies with more than 1,000 employees, with the rest from smaller organizations. 80\% of participants worked in the US. Participants held a variety of positions within their organizations (Table \ref{tab:roles}) and came from a diverse set of sectors (Table \ref{tab:sects}). For the categories in \ref{tab:roles}, “External Consultant” includes outside auditors and advisors for various issues surrounding system bias and fairness. The “Leadership Role” category is used to describe participants overseeing the work of larger teams (e.g. “Director/Head of X”). “Tech Contributor” is an umbrella term for roles such as software engineers, engineering managers, and data scientists.

\begin{table}
    \caption{Participant Roles}
    \label{tab:roles}
    \begin{tabular}{ll}
    \toprule
    Role                                 & Count \\
    \midrule
    External Consultant (EC)             & 4     \\
    Leadership Role (LR)                 & 9     \\
    Legal/Policy (LP)                    & 4     \\
    Product/Program/Project Manager (PM) & 7     \\
    Researcher (R)                       & 4     \\
    Tech Contributor (TC)                & 10    \\
    \bottomrule
    \end{tabular}
\end{table}


\begin{table}
    \caption{Participant Sectors}
    \label{tab:sects}
    \begin{tabular}{lll}
    \toprule
    Sector & Participant IDs & Count \\
    \midrule
    Ad Tech & LP1, LP2 & 2 \\
    Finance & LR2, LR3, TC3, TC4 & 4 \\
    Healthcare & EC1, LR8, PM5, TC7 & 4 \\
    Hiring/HR & LR1, LR4, LR7, PM7, TC10 & 5 \\
    \begin{tabular}[c]{@{}l@{}}Multi-Sector\\\hspace{1mm}  Technology\end{tabular} & \begin{tabular}[c]{@{}l@{}}LR5, PM2, PM3, PM4,\\   \hspace{1mm}R1, R2, R3, TC2, TC6\end{tabular} & 9 \\
    Social & LP3, LP4, LR6, PM6, TC5, TC9 & 6 \\
    Telecom & LR9, PM1, TC1 & 3 \\
    Other & EC2, EC3, EC4, R4, TC8 & 5 \\
    \bottomrule
    \end{tabular}
\end{table}

We employed a variety of recruitment methods. Firstly, we identified individuals in the authors’ professional networks that would most likely have experience with attempting to implement algorithmic fairness techniques. These contacts were asked to participate themselves and were also encouraged to share our call for participants with relevant individuals in their own network. Similarly, we distributed an interest form and study description to various organizational mailing lists. In order to broaden the search beyond our established networks, we also carried out searches for various participant archetypes through the LinkedIn Recruiter service \cite{LinkedIn}. As suggested by \citet{maramwidze-merrisonInnovative2016}, LinkedIn can be a prime mechanism for identifying and contacting specialists that would otherwise not be accessible to the researcher. By using a Recruiter account, we were able to send requests for participation to individuals with greater than two degrees of separation from the account-holder. Given the typical use case of this medium, we clearly stated that we were not reaching out with an employment opportunity. In addition to these methods, we also conducted searches for news articles and organizational publications on topics related to algorithmic fairness and bias auditing in order to identify teams and authors to directly reach out to. Finally, following each interview we asked participants to refer any relevant contacts. With all of these recruitment methods, we attempted to sample participants from industries with varying degrees of regulation, as we expected this to be an important axis of analysis based on the existing literature.

Participation in the interviews involved one 60-75 minute video call. In some cases, we agreed to have multiple interviewees from the same organization participate in the same call. Before the call, we shared our consent and confidentiality practices with participants and then obtained verbal consent to record before starting the interview. We transcribed the audio recordings, and deleted them after transcription.  We redacted all findings to maintain the anonymity of participants and their institutions. 

Since the term “demographic data” may mean different things to different people, we provided the following, rather broad, definition at the beginning of the interviews.

\begin{quote}
    Demographic data includes most regulated categories (e.g., “protected classes” and “sensitive attributes” like sex, race, national origin) as well as other less-protected classes (e.g., socioeconomic class and geography). Demographic data can even include data that might be used as proxies for these variables, such as likes or dislikes.
\end{quote}

The interview questions focused on asking participants to walk us through examples of times when they had participated in successful or unsuccessful attempts to use demographic data for bias assessments. We probed for details about what data they wanted to use and what their intended purpose was. We asked about the availability or obtainability of the desired data. And we asked followup questions about what types of approvals or interactions with other teams were required for the collection or usage of this type of data. 

In cases where the participant was in a legal/policy role, the questions were similar to the above, but framed from the vantage point of someone reviewing a request for data access or usage. In these interviews, we also asked additional questions about what legal or other risks they take into account when making decisions. In cases where participants were identified based on press reporting or organizational publications on algorithmic fairness issues, we sometimes included more tailored questions about the content of the publication. 

After hearing about more specific cases, we also asked for more general reflections from participants on the trade-offs involved in this type of bias assessment work --- e.g., “How do you balance user privacy concerns with the need for demographic data?” In addition, we heard some insightful reflections in response to more forward-looking questions such as “In your ideal world, what would bias assessment look like for you?”

To analyze the interviews, we used open and closed coding in MaxQDA. Open-coding was used to parse and summarize an initial selection of 25\% of the interviews. Open codes were then iteratively grouped and synthesized using thematic networks \cite{attride-stirlingThematic2001}, with themes organized around either the challenges with or constraints on demographic data use. These themes and their subthemes then informed a set of closed codes that were applied and expanded upon over the corpus of transcripts.

\section{Results Overview} \label{sec:over}

Before delving into emergent themes, it is important to first note some of the broad trends we heard concerning the availability of demographic data. Almost every participant described access to demographic data as a significant barrier to implementing various fairness techniques, a result that mirrors the responses from \citet{holsteinImproving2019}. In terms of accessible data, virtually all companies that collected or commissioned their own data had access to gender and age and frequently used that data to detect bias. These categories of demographics data were regarded as entailing much less sensitivity and privacy risk than other categories. Outside of employment, healthcare, and the occasional financial service contexts, few practitioners had direct access to data on race, so most did not try to assess their systems and products for racial bias. When attempts were made, it was generally through proxies or inferred values, but such methods were rarely deployed in practice (conforming to previous findings by \citet{holsteinImproving2019} for demographic attributes more generally). Finally, practitioners in business-to-business (B2B) companies and external consultants generally had very limited access to sensitive data of any kind.

We organize the themes uncovered through our analysis into two types. The first type are the regulatory and organizational constraints on demographic data procurement and use, which are generally maintained by a network of actors outside the practitioner’s team (e.g., legal and compliance teams, policy teams, company leadership, external auditors, and regulators). The second type are concerns surrounding demographic data procurement and use that the practitioners themselves surfaced or encountered during the course of their work. Having made this distinction, however, it is important to note that it was not uncommon for the constraints in the first category to inform the concerns of practitioners or for the concerns of practitioners to feed into organizational policy.

\section{Legal and Organizational Constraints on Demographic Data Procurement and Use} \label{sec:constraints}

Looking first to the network of constraints surrounding the procurement and use of demographic data, we discuss three major factors at play. The first, and perhaps most obvious, are the various regimes of privacy laws and policies. Second, we consider the role of anti-discrimination laws and policies in determining whether demographic data is collected. Finally, we consider a suite of potential restrictions stemming from organizational priorities and practices.

\subsection{Privacy Laws and Policies} \label{sec:priv}
GDPR-related notions like data-subject consent and the Privacy By Design standard of only collecting required data seemed to primarily guide legal, policy, and privacy teams’ handling of data requests, despite most participants being based in the U.S. As described by LP1, a legal practitioner in the Ad Tech space: 

\begin{quote}
    “The general framework is that the rules apply to data about people. And then, one axis on there is ‘what does it mean that the data is about a person.’ [T]he old school way of looking at that was [whether the data was] identifiable, which meant it had their name or contact information, email address. The newer GDPR-like way to look at it is that even pseudonymous data --- [i.e.,]  I'm  not [First Name, Last Name], I'm ABC789 (or my computer or my browser is) --- even pseudonymous data is regulated.”
\end{quote}

\subsubsection{Consent and Special Category Data.}
For this work we left our discussions of “demographic data” open to any membership category of data relevant to our participants. Generally speaking, however, practitioners pointed to the categories defined by law as protected or sensitive. An important distinction that emerged for organizations with operations in Europe was between categories classified as “special” by GDPR and those that are not. Most notably, gender, age, and region, three categories not considered “special,” were available to most of the practitioners we spoke to, whereas race, a “special” category, was not. This is especially significant because race was seen to be a very salient axis of analysis for every domain, yet attempting to procure data on race was rare. The absence of this data was keenly felt by practitioners attempting to show the failure modes of various systems, as depicted by TC9, a product manager in the social media domain: “Every time we do one of these assessments, people ask, ‘Where's race?’”

Though few participants saw the requirements set by GDPR around special category data as entirely prohibitive, many discussed how the general practice is to just avoid the risks of wide-scale special category data collection altogether. In the rare cases where special category data was procurable at a large scale, it was either through government datasets, through open-sourced data with self-identification, or by following explicitly applicable regulation and standards. There were also some reported cases where racial data was collected for user studies and experimentation, but this was done with informed consent and compensation, as well as strict legal review.  The only domain in which we saw the attempted collection of racial data outside of the U.S. was for hiring and HR. \footnote{Hiring and HR carve-outs for special category data in regional GDPR implementations tend to defer to employment law in determining what data can be collected and for what reasons, making it potentially less fraught than for other domains. See e.g., the German Federal Data Protection Act Section 26 (3) \cite{bdsg}}

\subsubsection{Don't Collect More Than What You Need.}
Beyond the legal risks incurred by potential GDPR violations, we also heard from practitioners that it is “the combination of GDPR and the ethos that GDPR lives in, which is an ethos of tech fear and nervousness and lack of trust” (PM3) that can lead to inaction. If public trust in data use is already low, the Privacy By Design standard of not collecting more data than you need can very easily take precedence over the largely experimental process of ensuring product fairness. Describing an experience voiced to us by multiple participants, TC5 talked of how “for the collection of new data, there's deep conversations with policy [and] legal, and then we need to loop in external stakeholders [to ask], ‘does the need to be able to measure this seem to sufficiently balance out the user concerns for privacy?’” In many of our interviews, we heard practitioners recall cases where data requests would be stifled by this question of necessity. 

Interestingly, practitioners reported that inferring demographic data is similarly frowned upon. One potential reason for this is the increasing prevalence of data access request rights through regulation such as GDPR and the California Consumer Privacy Act. LP4 discussed how with “data access requests, it’s still an open question under certain laws whether inferred data is \textit{user} data and [thus] needs to be shared or not. And so [a] concern around that [is], ‘are people aware that this might be happening?’” In this way, the potential of forced revelation of inferred data may be encouraging corporations to be more conservative about what they infer. 

\subsubsection{Data Sharing.}

One final difficulty often presented by data privacy regulation in this space is the inaccessibility of demographic data for outside auditors or consultants as well as for business-to-business vendors of algorithmic tools. Though we did hear of cases where external auditors were brought into an organization and given pseudonymized access to data, we also heard of many cases where auditors, consultants, and vendors had to build their models or provide recommendations without ever seeing the data. For practitioners in this kind of situation, some spoke of the importance of publicly available datasets drawn from domains similar to the organizations they were assisting. By combining this external data with API access or knowledge of the infrastructures in use, these practitioners noted that they could point to specific, high-risk failure modes without needing to risk any legal privacy violations. In one unique case, a practitioner opted to forego the use of public/private data altogether, instead making the case for “pathways” of discrimination in conceptual Structural Causal Models (SCMs) built in concert with subject matter experts and other employees from the organization they were assisting. The SCMs was then used to inform the client of potential sources of bias in their data, despite not taking in any of that data.

\subsubsection{Implications.}

While there are many other motivations for some companies to not actively collect “special category data,” GDPR has raised the bar for private organizations that wish to use this data for the purposes of bias detection or fairness assessments. Even if GDPR does not prohibit the collection of sensitive attributes for bias detection (as outlined in recent guidance from the UK Information Commissioner's Office \cite{icoai}), increasingly protective privacy regulation has empowered privacy and legal teams to err on the side of caution when it comes to data sensitivity. This is likely why many practitioners discussed bias detection efforts being denied if they required new data. As such, fairness practitioners either are pushed to inaction or have to be creative in how they assess and address bias, as we explore in Section \ref{sec:concerns} below.


\subsection{Anti-Discrimination Laws and Policies} \label{sec:anti}
Most often, the closest legal and policy teams come to thinking about fairness is through anti-discrimination policy. For this section, we first consider a few domains where we saw clear anti-discrimination standards interact with the collection and use of demographic data, and then move to looking at how this interaction plays out when liability for discrimination is less clear.

\subsubsection{Finance.} The few practitioners we spoke with from the Finance sector were quick to discuss the differences between mortgage products and credit-based products in the U.S. For the former, lenders are required by law to collect demographic attribute information, while for the latter they are nearly barred from doing so. In both of these cases, however, financial institutions are held to anti-discrimination standards by various watch-dog institutions, such as the Consumer Finance Protection Bureau (CFPB). For products where the data required to ensure compliance is sparse or cannot be directly collected, these participants discussed how their institutions have followed the CFPB's lead in inferring racial categories through the Bayesian Improved Surname Geocoding (BISG) \cite{elliott2009using} method. Even so, they reported never having direct access to these attributes themselves, inferred or otherwise. Instead, compliance teams with special data access will test models after they have been produced, sending them back if they are found to be discriminatory. When discussing this arrangement, TC4 suggested that "they kind of want the whole testing function to be a black box," in part so that no intentional circumvention of compliance tests can occur. This can have the unfortunate effect, however, of leaving engineering teams in the dark about the impacts of their models. 

For some types of discrimination, even clear regulation does not necessarily lead to measurement and mitigation. On the issue of regulations prohibiting financial service discrimination based on sexuality, LR3 noted that, “in practice, the regulators understand that data for certain classes are not easy to obtain and so they don't make any enforcement. [...] So it's one of those things where when the rubber hits the road, you have to figure out, is this actually something you can do?”  Even for classes of data that are possibly easier to obtain, financial institutions can still be resistant to their collection depending on how anti-discrimination is enforced.

\subsubsection{Human Resources.} In the Hiring/HR domain, participants referenced clearer standards that they defer to. In the U.S. context, practitioners mentioned how the demographic categories on EEO-1 forms (gender and race/ethnicity) were often made available to them such that they could ensure their models’ adherence to the "80\% rule". \footnote{The EEOC's adverse impact rule of thumb, where if the selection rate for one group is four-fifths that of another it may point to deeper issues \cite{us1979questions}.} Difficulties can arise when the organization’s standard practices for bias and discrimination assessment go beyond the expectations, requirements, and norms for doing so in the country of business, however. When discussing doing business in a country outside their own, PM7 described how, “We talked about demographics and they were like, ‘This is not something that is important to us.’ [...] It becomes a real challenge to get customers even interested in why we do [model debiasing].” As a result of these types of concerns, participants discussed bringing on employment anti-discrimination experts specializing in the various regions they operate in. In terms of how procured demographic data often gets used, teams we talked with that did model debiasing relied on specially collected and curated datasets, generally from client organizations, that have demographic attributes of reliable accuracy. Typically, this type of high-quality data is only used for the purposes of reducing discriminatory effects before deployment, as there are too many practical and legal constraints on including demographic data in deployment.

\subsubsection{Healthcare.} Healthcare was the third and final domain where we spoke to participants that pointed to concrete anti-discrimination policies. Though not necessarily codified in law, professionals from one organization noted that they try to strictly adhere to the Institute of Medicine’s Six Dimensions of Health Care Quality \cite{instituteCrossing2001}, one of which is health equity. This established what they saw as clear alignment between the goals of their organization and the push for anti-discrimination in algorithmic systems. Furthermore, practitioners reported that demographic data is often required for treatment purposes anyways, so obtaining it for fairness assessments is generally not an issue. Interestingly, this was also the main domain in which practitioners saw the potential of algorithmic fairness as extending beyond interventions to the algorithm or product. While in other domains societal bias could sometimes be seen as beyond the scope of the practitioners’ organization, practitioners in healthcare reported using their trained models to inform interventions anywhere in the patient treatment pipeline. In other words, uncovered unfairness did not necessarily have to stem from institutional practices or technical models to warrant addressing.

\subsubsection{Less Regulated Domains.} In cases outside of these domains, we saw an abundance of uncertainty around anti-discrimination policy that seemed to lead to risk-aversion and inaction. One of the biggest constraints participants described was that even without clear anti-discrimination policies, there were still fears that procuring demographic data and using it to uncover discrimination without a clear plan to mitigate it --- either because doing so might not align with the organization’s priorities or because there are no clear answers for how to resolve the issues --- would open the door to legal liability. As PM4 described, “I think there's always this sort of balancing test that the lawyers are applying, which is weighing whether the benefits of being proactive about testing for bias outweigh the risks associated with this leaking or this becoming scooped up in some sort of litigation.” Participants also expressed the sentiment that unless an organization is distinguishing itself by its commitment to fairness (however defined) or directly responding to outside pressure, it is often easier to just not collect the data and not call attention to any possible algorithmic discrimination.

\subsubsection{Implications.}
As described in a congressional testimony given by the GAO \cite{williams2008fair}, financial institutions are not confident they will not be held liable for disparities uncovered through the collection of demographic data, incentivizing inaction. In other domains as well, the spectre of increased discrimination liability looms large when it comes to demographic data collection. So long as there is a dearth of applicable legal precedent, we are relying on self-regulation in a context where often the safest move is to simply not act.

Drawing on the strategies seen in the healthcare domain and as has been suggested elsewhere \cite{barabasInterventions2018,andrusJust2019}, algorithmic fairness assessments can be used as a type of diagnostic tool for the whole network of actors and interactions, identifying salient disparities in intake, treatment, and outcome and pointing to possible causes for these disparities. This approach, however, goes well beyond what is currently required by anti-discrimination law (in cases where requirements even exist) and will likely require a shift in political will to meaningfully take root in other domains.

\subsection{Organizational Priorities and Structures} \label{sec:org}
As in any corporate setting, practitioners working on algorithmic fairness face constraints set by organizational structures, risk aversion, and the drive for profit. Though some practitioners described provisions for fairness practitioners to do their work unbeholden to the short-term goals of other product teams, many still described constraints imposed by organizational priorities and practices. 

\subsubsection{Fairness versus KPIs.} Starting with issues arising from tenuous fairness commitments, multiple practitioners mentioned having to justify their bias analyses and calls for more or better data with expected improvements to key performance indicators (KPIs). In these cases, it was not uncommon for practitioners to find that their proposed interventions were orthogonal to certain KPIs, such that they had to make qualitative arguments about how different demographic groups, and in turn the company, would be better served by their interventions. Practitioners who had tried this approach divulged that if it failed, the work could just be abandoned. A similar constraint came from practitioners reporting that at times the cost of gathering more detailed or more representative data simply came with too high of a price tag and too long of a lead time to be deemed worthwhile.

\subsubsection{Organizational Momentum.} Another significant concern we heard was that often fairness and responsibility teams do not have clear pathways for pushing recommendations into practice. Other teams’ existing processes around data use and procurement are generally already well-established, and so it can be very difficult to figure out both how to plug in operationally as well as how much those processes can be overwritten or overloaded in the pursuit of fairness.

\subsubsection{Public Relations (PR) Risk.} A final constraint on this front is the need to avoid bad PR stemming from demographic data collection. Practitioners expressed concern that efforts to collect demographic data are likely to be seen as an extension of the frequently reported trends of data misappropriation and data misuse (See e.g., \cite{seneviratneUgly}). Specifically addressing the issue of racial data collection at their organization, PM2 discussed how, from the public’s perspective, such data collection would be viewed with suspicion: “[It's] just adding one more thing. [The public might say,] 'now you're asking me if I'm White or Black, why would you need that information?' Right? ‘You're already in trouble with data. How can I trust you with this?’ So I think it was an optics thing, like, we can't even ask that. It's kind of off the table.” The calculus can change dramatically, however, in cases where there is already a lot of public attention around a company’s biased services. As described by EC3, “public interest and attention and focus -- it actually puts these companies at a tipping point where they can't just do nothing, because even though they can say they're complying with all the laws, clearly the public doesn't feel like that's adequate.” In cases such as these, participants reported that their companies would start up limited and measured data collection processes to address, at least in part, the source of public concern.

\subsubsection{Implications.}
It is important to note that many of these constraints seem to exist on a spectrum of severity, largely dictated by the amount of buy-in from both leadership and the organization’s clientele to ensuring algorithmic fairness. As organizational practices around fairness shift and solidify, fairness practitioners will likely need to play a central role in making the argument for collecting demographic data despite these organizational risks \cite{rakovaWhere2020}.

Tying into Section \ref{sec:anti}, the PR risk of the appearance of discrimination is also likely to translate into some legal risk. Though this risk was rarely discussed by participants, past work by the authors has outlined some of the ways the attempts at algorithmic fairness might be litigated against for being discriminatory under certain legal interpretations of ant-discrimination law \cite{xiang2020reconciling}. It remains to be seen, however, how these types of cases will play out in court.

\section{Fairness Practitioners' Concerns}
\label{sec:concerns}

Beyond business, legal, and reputational pressures from other parts of the organization, we see a number of sources of caution surrounding the procurement and subsequent use of demographic data on the fairness practitioner side.  

\subsection{Concerns Around Self-Reporting} \label{sec:self}
Regarding the actual process of procuring demographic data, practitioners reported a number of concerns around the effectiveness of current collection methods.

\subsubsection{Frequency and Accuracy of Responses.} The most commonly reported concern about self-reported demographic data was that it is often unreliable or incomplete. In many cases, there are not strong incentives for individuals to respond to requests for their demographic data. Outside of government surveys and forms, participants noted that organizations often need to incentivize individuals to provide accurate data. \footnote{This is by no means a new problem, social scientists have long been studying how to increase survey response rates and the biases associated with non-responses \cite{marquis1986response}.} In the case of the American financial industry, one participant discussed how although they are permitted by federal law to collect demographic data through optional surveys, the response rates are incredibly low. While it might have been possible to improve the response rate by making the intended use for the data clearer \cite{crottyRevised2020}, this participant noted that more reliably collecting demographic data could actually increase the liability of the institution, as discussed in Section \ref{sec:anti}. As such, the sparse survey data was only used for validating data that had been compiled and inferred via other sources. 

There are many reasons why a survey might get a low response rate, but the most frequently discussed reason was the matter of trust. LR6 discussed how when their team was working on improving search diversity for their social media platform, they surveyed users asking, “would you want to give us more demographic information?” LR6 described how “the consensus was, ‘no, [if you collect this data,] then what are you not showing me? [...] Who are you to decide what I would like just because I am this person?’” 

Within the healthcare domain, practitioners were acutely aware of the types of difficulties associated with accurate data collection, as they collect demographic data for treatment analysis and to enhance their ability to provide care. As LR8 described, “We think about data collection as data donation, it's like donating blood. It takes effort, someone's got to do it, right? We don't have [an] automated way to collect all this information, so we have to think carefully about the most effective and efficient ways to collect [...] most of the data that we collect.” 

\subsubsection{Collection Procedure Difficulties.} On top of data reliability concerns, how organizations would even go about the process of collecting demographic data is unclear. EC4, an expert who assists platform companies trying to deal with potential bias or discrimination, discussed how an organization might “want to do a disparate impact [analysis] on gender” when “[they] don't have gender." EC4 asked, "Are [they] going to throw up a [pop-up] and say, ‘Hey, tell us about your gender identity,’ to their whole user base? There's some cost to doing that.” Due to the range of potential public responses and PR risks as described in Section \ref{sec:org}, the reputational cost could even exceed the expense of designing and deploying such a pop-up. So, as EC4 put it, whether “any company is ever going to do that to a user base of millions of people that are already on their service” remains an open question. 

\subsubsection{Flexible Categories.} Furthermore, some practitioners discussed how even reliably collected demographic data can lead to issues down the line. Most often, demographic data is self-reported by selecting only one answer from a set of predetermined options, which may or may not include an “Other” category. What individuals signify by selecting one of these options, however, is not necessarily consistent. TC9 outlined a number of the potential concerns that can arise here surrounding gender:
\begin{quote}
    “And I think one thing that I'm very conscious of is that anything can be misinterpreted. [...] So for example, we have four gender options, ‘male,’ ‘female,’ ‘unknown,’ and ‘custom gender’, where ‘custom gender’ is an opt-in, non-binary gender that a user can add a specific string for. ‘Unknown’ is you haven't provided it or you've explicitly said you don't want to tell us your gender. It's a whole bunch of things. And even though this is not inferred data, I'm still very, very careful whenever I talk about gender analyses, to be really clear about what exactly we're seeing here, because ‘custom gender’ does not mean non-binary, ‘unknown’ gender does not mean the person opted into not giving it to us.”
\end{quote}
Similarly, other practitioners described exercising caution around the demographic data they do have access to, acknowledging the myriad identities that can fit within a single box on a form.

\subsubsection{Implications.}
At a high level, when data subjects are wary of the reasons behind or potential outcomes of demographic data collection, it is reasonable to expect that the response rate and response accuracy are likely to be very poor. While it is likely to be difficult to transfer the healthcare style of thinking to other domains, posing data collection for anti-discrimination and fairness assessments as something akin to “data donation” might be a useful framing for increasing buy-in and response rates.

\subsection{Concerns around Proxies and Inference} \label{sec:prox}
Where practitioners’ concerns around the collection of demographic data mirrored some of the well-established pitfalls and risks of survey design, we see a rather different set of concerns around the use of proxies and inferred demographic data. For sensitive demographic features that are unlikely to be approved for collection by various oversight teams (often for reasons given in Section \ref{sec:constraints}), we sometimes see practitioners infer attributes directly using available data or look at other available categories of data that have an implicit correlation with attributes of interest.

\subsubsection{Prevalence of Proxies and Inferred Categories.} When asked whether, and if so how, inferred demographics were used in any of their fairness analyses, we heard a wide range of responses from practitioners. Participants from U.S. financial institutions noted that they follow the standard set by the Consumer Financial Protection Bureau in using Bayesian Improved Surname Geocoding (BISG) to infer individual race \cite{bureau2014using}. Participants from domains where demographic data collection is mandated, such as the HR/Hiring space, on the other hand, balked at the question, as exemplified by the response of LR4: “We both A) don't have a reason to do that in the line of what we're doing, B) would be deeply opposed to doing that in this context. [...] Fascinating, I would want to know who else would do that in what context and why you would ever do that?” 

Responses from participants in other domains were situated between these two extremes. Oftentimes inference is the only available option for conducting discrimination or bias assessments, but there are few standardized practices for doing so. Of the participants that discussed using demographic inference, it was largely for content data, such as in skin-tone classification for images or author gender prediction for pieces of text. Inferring demographics in other cases, while perhaps feasible, was seen as introducing privacy risks as well as dignity concerns to any fairness evaluations conducted with them. These concerns were especially heightened around sensitive attributes. As described by TC5, “In contexts of race or sexual orientation, or other really sensitive groups, we don't view inferring attributes without fully informed user consent as a path forward. We think that all of the concerns are around whether [our organization] has the data. And so, if we were to try to pursue [acquiring the data], we would need to do it explicitly and, you know, with fully informed consent.” Other practitioners suggested that when inference is done at the group level (i.e. ”X percent of this dataset is women”), that might resolve some of the consent and dignity concerns of classifying individuals. 

\subsubsection{When Inferred Categories are Preferred.} Although inferred demographics are generally used in contexts where self-reported demographics are not available, an interesting dilemma arises in cases where inferred demographics are more suitable for the task. For instance, EC3 discussed how on one of the systems they worked on, there were potential issues based on how users perceived other users’ races: “So you're doing perceived race. Then the question is, so if I'm a tech company and I want to study this racial experience gap, right, should I go around guessing everybody's races?” While having labelers or an algorithm guess someone’s race is perhaps the best way to operationalize the concept of perceived race, EC3 and others made sure to point out that it opens the door to subsequent backlash given the sensitivity of these labels.

\subsubsection{Treating Proxies as a Weak Signal} A unique set of concerns arises for proxies that are not explicitly treated as such. In some cases, when a salient attribute is inaccessible, practitioners use other available attributes to obtain some signal about potential discrimination or unfairness surrounding that more important attribute (e.g., using a subscription tier-level to point towards potential problems across socio-economic status). In cases such as these, practitioners were very clear that these proxies are not meant to be treated as the attribute itself: TC9 explained, “We're very careful to not draw further conclusions, because we don't want to make assumptions or directly infer [the attribute of interest], basically.”  Instead these attributes are treated as a very rough signal, where if you notice large disparities, you should treat it as a call for deeper inspection.

\subsubsection{Implications.} Although the algorithmic fairness literature has frequently called out the dangers of using proxies for salient attributes or target variables in fairness assessments \cite{barocasBig2016,corbett-daviesMeasure2018a,jacobsMeasurement2019}, this is often the only option practitioners have to make a quantitative argument for escalation and deeper analysis. The real risk that we see arising with this practice is when/if the use of the proxy becomes sufficiently normalized such that practitioners no longer reflect on its inadequacies. Proxies should not just be plugged into standard bias mitigation techniques without first clearly outlining how they constitute a "measurement model" of the desired attribute \cite{jacobsMeasurement2019}. Using them to uncover egregious discrepancies, however, might inspire interrogation of the full decision-making pipeline. This in turn could reveal more about the conditions of unfairness than would a technical tweak to the algorithm.


\subsection{Relevance of Demographic Categories} \label{sec:rel}

Most types of fairness analysis require practitioners to identify and focus on discrete groups, but there is often discomfort and uncertainty around what groups warrant attention and how those groups should be defined. These concerns were seen to be especially salient within organizations and teams that had to build and maintain products and services for regions outside of their own. 

\subsubsection{Deficient Standards.} Looking first to the definition of demographic categories, numerous practitioners discussed the inadequacy of the standard categories. This was especially a concern in industries where demographic data is collected according to governmental standards. PM7, a product manager in the hiring/HR domain, pointed to how they were actively “finding ways to gather more data outside of the employment law data,” as “the categories were defined back in the sixties and are challenging these days. [...] You only get six boxes for your race. There's two boxes for gender. You know, it just doesn't work for a lot of people.” For other domains, the established categories were sometimes just used out of tradition or because it would be hard to change them.

\subsubsection{Difficulties with Granular Representations.} A common response to this problem was to try to make the categories representative, but practitioners were generally unsure of where to draw the line. R3, who was tasked with expanding the demographic categories and variables of concern used by their organization, said of this issue: “You can go on and on and create a list that's like 20 or even a hundred long, but we need to set a limit, [...] like what's the reasonable limit on how many categories you want to put in such a question and what we want to recommend internally to like product teams to at least sample across.” A common concern here is that you start to lose statistical significance as the groups get smaller. Concerning asking more granular demographic questions, a practitioner from the Healthcare domain (LR8) asked, “how should we capture [...] relatively low frequency answers to those questions? [...] All of a sudden you become an N of 1 and you don't fit into any category and we can't figure out how to help you. Whereas, if we could have lumped you into another category, might we have been able to do better?”

\subsubsection{What Categories Should Be Focused on?} Whether or not the demographic categories are expanded before data collection, there still remains the question of what the most relevant categories or combinations of categories are to consider for evaluation and deeper inspection. PM4 said of this problem, “[this] step of [asking], ‘what demographic slices are most relevant,’ that's very difficult to get because almost nobody feels like they have the authority to say, ‘these are the ones that matter.’” While it might not require much extra effort or resourcing to assess bias across all available attributes, practitioners' teams were also generally responsible for informing algorithmic, design, and policy changes. On top of this, several practitioners recounted needing to find bias issues with groups at the intersection of multiple attributes as well \cite{crenshaw1989demarginalizing,hoffmannWhere2019}. Considering all possible intersections of demographic attributes is generally not practical for a small team (e.g., if you have five attributes, each with four options, then there are $4^5 = 1,024$ total groups), so some practitioners had used written feedback from clients and users to help identify which groups their systems were not working well for.

Arguably the most straightforward approach to this problem is to just defer to legal or policy requirements around what groups to consider, but they often do not map onto practitioners' views of what is ethically or culturally salient. R1 notes that this can become a significant impediment to making systems fairer in practice: “I'd say that the main conflict is where we'll do some testing on a product and reveal issues. And we'll say ‘look, this, it doesn't perform on this group.’ Or, ‘it makes these failures.’ And then the policy team will say [something] like, ‘well, we don't have a specific policy about this, so why are you testing for it?’” When these conflicts arise between the types of categories practitioners think are most salient and the categories that their organizations expect them to assess bias for, a substantial minority of participants reported having to just shelve their work on those categories.

\subsubsection{Regional Differences.} An additional concern reported by practitioners was that their companies’ awareness of what categories are legally salient is often based on the laws of the region in which the company is located. Concerning the general focus on “race, ethnicity, gender, and maybe sexual orientation” at their organization, PM4 remarked, “I often think that the way we approach this is very U.S. centric, that our understanding of what demographic characteristics would be relevant in other countries and cultures is pretty rudimentary.” Some organizations try to ameliorate this concern by introducing subject matter experts with understandings of regional and cultural standards, but this is often only if there are explicit anti-discrimination regimes they have to work within. In reference to their algorithmic HR support tools, LR1 said, “The main problem is local employment law. We don't have knowledge of local employment law, and every country is different. And so we always work with what we call a HR business partner, who is able to bring in labor experts if needed.”

\subsubsection{Convenience.} Finally, when the relevant data is just not available for testing, it may lead to a default focus on whatever category data is on hand. When discussing ensuring dataset diversity beyond gender, PM1 stated that “it's a complex problem because we are not allowed to collect such [demographic] data. So that's why gender is prioritized, because we can do something about that.”  Similar to the reliance on loose proxies, this issue of focusing on categories due to availability and not social salience or public demand was a common concern relayed to us by participants. 

\subsubsection{Implications.} Adhering to inapt categorization schemas can be an impediment to meaningfully understanding the types of bias present, as the demographic categories used can lead to dramatically disparate analyses and conclusions \cite{howellWhat2017, hannaCritical2020}. Tackling this risk will likely require earnest attempts to grapple with specific issues in limited contexts, as the most salient forms of possible discrimination are likely to be missed with a "one-size-fits-all" approach that prioritizes scalabilty.


\subsection{Ensuring Trustworthiness and Alignment} \label{sec:align}
Though there is seemingly an alignment between model developer and data subject when implementing fairness interventions, practitioners discussed how this relationship can be difficult to bear out in practice due to both low public trust in technology companies and weak alignment between company and community goals.

\subsubsection{Generating Well-Founded Trust.} On the issue of trust, some practitioners in less-regulated industries expressed concerns about whether their organizations could take the steps to meaningfully garner public trust around demographic data collection. Often citing recent PR-cycles around data misuse and mishandling (e.g. \cite{lapowskyHow2019}), practitioners expressed concern that users and clients simply do not believe their data is going to be used in a way they have control over or for their benefit. We even heard from a few practitioners that they were not confident themselves that data collected for bias, fairness, or discrimination assessments would be sufficiently sheltered from other uses. One potential path forward here came from external auditors and consultants. EC4 noted that such external entities could require organizations to make enforceable, public commitments to narrowly use specially collected data only for fairness and anti-discrimination purposes in order to surmount concerns around misuse.

Some participants did, however, express hopes that their organizations might be moving in a direction of making clearer what data they would like to collect and why. A key element of trust-building, as one practitioner pointed out, is that the relationship needs to go both ways --- users and clients provide the data needed to assess system performance, and then organizations follow through on their commitments, providing at least a summary of the results back to users and clients. As described by LR9, “trust is a word that is almost meaningless. Trustworthy is something else, and much more weighty. [...] Trustworth[iness] is showing capacity on both sides of the trust dynamic. [Saying] someone is trustworthy means that they can receive information and treat it respectfully. Trust alone is...toothless.” There are likely to be many organizational and legal barriers to this type of trust-building exercise, but some practitioners felt that they might need to start exploring other ways of addressing fairness if the sensitive data they currently need cannot be collected in a trustworthy, consensual way.

\subsubsection{Alignment with Data Subjects.} Building on the notions of trustworthiness and consent, a few practitioners reported trying to reach a higher level of alignment with their data subjects. Concerning the use of demographic data in their research, R2 broached the question, “if you recognize that this system is working poorly for some demographic, how do you rectify that in a way that meaningfully pays attention and gives voice to this group that it's not working for?” By showing a commitment to centering the perspectives and needs of the groups that are either failed by or directly harmed by algorithmic systems, R2 and a few other participants suggested that it is more likely individuals from those groups would be willing to share personal information, including demographic attributes.

\subsubsection{Implications.}
Reinforcing practitioner perspectives that the public is wary of corporate data handling, recent surveys have shown that Americans are both largely unclear on how their data gets used and are not convinced that it is used for their benefit \cite{auxier2019americans}. R2's question around giving voice to groups that systems do not work for points to one possible path forward in making systems more trustworthy and resolving this trust gap. By ensuring that data subjects not only have a say in what data gets collected, but are actually involved in the thinking about how the data should be used, it is both more likely that data will be employed in the interest of those it is about and that the fairness achieved will have material benefits for their lives \cite{greenAlgorithmic2020,dobbeBroader2018,mulliganThis2019,martinjr.Participatory2020,katellSituated2020, viljoenDemocratic2020}.

\subsection{Mitigation Uncertainty} \label{sec:miti}
If practitioners are uncertain about how effectively they will be able to use demographic data and make changes based upon it, they can be deterred from trying to overcome the barriers and concerns described in previous sections. 

\subsubsection{What Should Be Done About Uncovered Bias.} As discussed in Section \ref{sec:intro}, the algorithmic fairness literature provides a number of techniques for potentially mitigating detected bias, but practitioners noted that there are few applicable, touchstone examples of these techniques in practice. Moreover, the methods proposed often do not detail exactly what types of harms they are designed to mitigate, so it is not always clear which methods should be applied in which contexts. The only domains in which we saw practitioners proceed with confidence were those with relatively well-established definitions of what it means to make systems fair (e.g., Hiring/HR and Finance) and those where the solutions were not limited to algorithmic fixes (e.g. Healthcare). In the first case, precedent such as the 80\% rule (see Section \ref{sec:anti}) gave practitioners confidence that they could appropriately mitigate uncovered bias. In the latter case, a lack of clear technical solutions did not hamper efforts to detect bias since practitioners could report evidence of bias to relevant stakeholders and the necessary interventions could be made. As an example, LR8 discussed a case where a model showed that individuals who spoke English as a second language had worse predicted health outcomes associated with sepsis. Instead of pursuing a technical intervention, they found some of the issues stemmed from an absence of Spanish language materials on the infection. In this case, had the spoken-language data field been either not collected or thrown out of the model as a naive anti-discrimination policy might dictate, equitable interventions would have been inhibited.

\subsubsection{Uncertainty About Whether Proposed Interventions Will Be Adopted.} Fairness practitioners often want to go beyond just the minimum requirements of corporate policy, but it is not always clear what will be permitted. One element of this is that that, as described in Section \ref{sec:rel}, their work can simply run ahead of what the policy team has considered. When conducting fairness analyses, practitioners mentioned surfacing uncomfortable questions, such as, what their responsibility is for historical inequalities that impact various performance metrics. When organizations are not prepared to answer these questions, it can be difficult for practitioners to know how to (and whether to) proceed with demographic data collection and use. A related issue is that when there are established policies on fairness, they generally seek to ensure all parties are treated “the same.” Whether this is through equalizing performance metrics across groups, such as calibration or predictive parity, or by removing/ignoring demographic data entirely, practitioners worried that achieving equity might require more nuanced treatments of different groups. As argued by TC9, “I would like to make a little bit stronger of a stand on [being] anti-bias. Instead of just making sure our models aren't biased, [we should] potentially carve out a space for treating groups differently and treating the historically marginalized group with more care.” 

\subsubsection{Implications.} Though organizational policies around product fairness and anti-discrimination are becoming more common, our interviews revealed that product and research teams are still the ones being left to grapple with what constitutes algorithmic fairness. While these teams are equipped to answer some of the questions that arise, optimizing for fairness can and will surface larger questions about the role of corporations in creating a more just world, so the task cannot be left to these teams alone. Furthermore, in the cases we heard about where these teams were given explicit guidance about how to resolve these dilemmas, it was usually via policies that attempted to just treat every group the same. This strategy of trying to espouse a “view from nowhere” \cite{harawaySituated1988}, however, often just means taking the perspective of the majority or those with power (e.g. multiple participants discussed hate-speech classifiers that would always classify texts with the N-word as hate-speech, implying that the speaker is not Black). Before new demographic data collection efforts are taken up for the sake of algorithmic fairness, we recommend that organizations first come up with a clear goal for their fairness efforts and a plan for how to empower the relevant teams to achieve those goals.

\section{Discussion}

Having mapped out this dense web of challenges practitioners face when attempting to use demographic data in service of fairness goals, we do not believe that the next step should be to simply lower the barriers to collecting demographic data. To the contrary, many of the challenges participants raised highlight deep normative questions not only around how but also \textit{whether} demographic data should be collected and used by algorithm developers. While the main goal of this paper was to explore the contours of these normative questions from the practitioner perspective, we take the rest of this section to consider three elements of possible paths forward: clearer legal requirements around data protection and anti-discrimination, privacy-respecting algorithmic fairness strategies, and meaningful agency of data subjects.

Looking first at legal provisioning, using demographic data to assess and address algorithmic discrimination will require clear carve-outs in data protection regulation. This could entail establishing official third-party data holders and/or auditors, or it could simply mean opening the door for private organizations to do this work themselves. During the course of this interview project, we did see the UK Information Commissioner’s Office publish explicit guidance on how to legally use special category data to assess fairness in algorithmic systems \cite{icoai}, but it is not yet clear what impact this will have.
The approach of relaxing demographic data protections by itself, however, will likely leave many of the other constraints and practitioner concerns unaddressed. Further still, anti-discrimination law does not necessarily provide incentives for using demographic attributes to inform algorithmic decision-making \cite{xiang2020reconciling,harned2019stretching,tischbirek2020artificial,bentAlgorithmic2020}. As such, data protection carve-outs will likely need to coincide with updated protections around algorithmic discrimination. Looking to domains where demographic data collection is already legally provisioned (e.g. HR in the U.S.), we see that practitioner concerns do not dissolve so much as change shape (e.g., issues with the government categories), suggesting the importance of crafting this legislation in collaboration with practitioners.

Turning to algorithmic fairness strategies that maintain privacy, they are often pitched as a means of enabling data collection in low-trust environments, ensuring such data can only be used for assessing and addressing bias and discrimination. \citet{vealeFairer2017} recommend a number of approaches that enlist third-parties with higher trust to collect the data and conduct secure fairness analyses, fair model training, and knowledge sharing. Other proposed methods build on this third-party arrangement, ensuring cryptographic or differential privacy from the point of collection \cite{kilbertusBlind2018,jagielski2019differentially}. Decentralized approaches such as federated learning could have similar benefits as they would possibly eliminate the need for data sharing entirely, though methods for fairness analyses are less built out in this area \cite{kairouzAdvances2019}. While these methods can address some of the trust-related concerns discussed by practitioners, they can also exacerbate other challenges. By off-loading the responsibilities of collecting, handling, and processing this data, the third-party would also inherit some of the practitioners’ more socially-minded concerns, such as how to determine demographic saliency and how to meaningfully resolve detected unfairness. Depending on the nature of the third-party organization and the level of investment from the primary company, this deferral of responsibility could increase the likelihood that these questions just remain unaddressed. 

The final approach we consider is more firmly incorporating data subjects into the work around fairness. As a relevant and timely example of why this is important to do, throughout the course of this project we saw debates flare up around the use of racial data in COVID response plans \cite{butlerEU,reventlowData2020,singhCollecting2020,jamesRacebased}. Opponents argue that without a strong commitment to actually intervening on and reversing systemic health inequalities, generating such data is more likely to perpetuate harm (e.g., through the construction of various negative narratives blaming disadvantaged communities for health disparities \cite{singhCollecting2020}) than to address it \cite{onuohaWhen2020}. In a similar way, we might expect hasty, uncritical pushes for demographic data collection to leave the roots of discrimination unaddressed---possibly even covering deeper issues with the veneer of having reduced algorithmic bias. As such, we see more transparent, inclusionary practices surrounding demographic data and algorithmic fairness \cite{katellSituated2020,wongDemocratizing2020,martinjr.Participatory2020} as a path forward that can clearly address many of the issues outlined above. Though meaningfully taking steps to to keep data subjects informed and to give them a voice in how their data is used is likely to be difficult and costly, doing so would mitigate privacy risks and a majority of practitioners’ concerns. Data subject consent is a widely consistent legal basis for demographic data collection, and concerns around data reliability would be reduced as the collection process comes to reflect what LR8 referred to as “data donation.” Furthermore, by deeply engaging various groups in the collection and use of demographic data, answers to questions around what the salient demographics are as well as how to meaningfully address uncovered bias will be closer at hand.



\begin{acks}
We want to thank all the busy interviewees that took the time to talk with us about their work and all the wonderful members of the Partnership on AI community that helped us scope this project. 
\end{acks}

\bibliographystyle{ACM-Reference-Format}
\bibliography{DemoData}
\newpage










\end{document}